\PassOptionsToPackage{unicode}{hyperref}
\PassOptionsToPackage{hyphens}{url}
\PassOptionsToPackage{dvipsnames,svgnames,x11names}{xcolor}
\documentclass[
  9pt,
  twocolumn]{article}
\usepackage{xcolor}
\usepackage[margin=1.8cm]{geometry}
\usepackage{amsmath,amssymb}
\usepackage{iftex}
\ifPDFTeX
  \usepackage[T1]{fontenc}
  \usepackage[utf8]{inputenc}
  \usepackage{textcomp} 
\else 
  \usepackage{unicode-math} 
  \defaultfontfeatures{Scale=MatchLowercase}
  \defaultfontfeatures[\rmfamily]{Ligatures=TeX,Scale=1}
\fi
\usepackage{lmodern}
\ifPDFTeX\else
\fi
\IfFileExists{upquote.sty}{\usepackage{upquote}}{}
\IfFileExists{microtype.sty}{
  \usepackage[]{microtype}
  \UseMicrotypeSet[protrusion]{basicmath} 
}{}
\makeatletter
\@ifundefined{KOMAClassName}{
  \IfFileExists{parskip.sty}{%
    \usepackage{parskip}
  }{
    \setlength{\parindent}{0pt}
    \setlength{\parskip}{6pt plus 2pt minus 1pt}}
}{
  \KOMAoptions{parskip=half}}
\makeatother
\usepackage{longtable,booktabs,array}
\usepackage{caption}
\usepackage{calc} 
\usepackage{etoolbox}
\makeatletter
\patchcmd\longtable{\par}{\if@noskipsec\mbox{}\fi\par}{}{}
\makeatother
\IfFileExists{footnotehyper.sty}{\usepackage{footnotehyper}}{\usepackage{footnote}}
\makesavenoteenv{longtable}
\usepackage{graphicx}
\makeatletter
\newsavebox\pandoc@box
\newcommand*\pandocbounded[1]{
  \sbox\pandoc@box{#1}%
  \Gscale@div\@tempa{\textheight}{\dimexpr\ht\pandoc@box+\dp\pandoc@box\relax}%
  \Gscale@div\@tempb{\linewidth}{\wd\pandoc@box}%
  \ifdim\@tempb\p@<\@tempa\p@\let\@tempa\@tempb\fi
  \ifdim\@tempa\p@<\p@\scalebox{\@tempa}{\usebox\pandoc@box}%
  \else\usebox{\pandoc@box}%
  \fi%
}
\def\fps@figure{htbp}
\makeatother
\NewDocumentCommand\citeproctext{}{}

\makeatletter
 \let\@cite@ofmt\@firstofone
 \def\@biblabel#1{}
 \def\@cite#1#2{{#1\if@tempswa , #2\fi}}
\makeatother
\newlength{\cslhangindent}
\newlength{\csllabelwidth}
\newenvironment{CSLReferences}[2] 
 {\begin{list}{}{%
  \setlength{\itemindent}{0pt}
  \setlength{\leftmargin}{0pt}
  \setlength{\parsep}{0pt}
  \ifodd #1
   \setlength{\leftmargin}{\cslhangindent}
   \setlength{\itemindent}{-1\cslhangindent}
  \fi
  \setlength{\itemsep}{#2\baselineskip}}}
 {\end{list}}
\usepackage{calc}

\newcommand{\CSLLeftMargin}[1]{\parbox[t]{\csllabelwidth}{\strut#1\strut}}
\newcommand{\CSLRightInline}[1]{\parbox[t]{\linewidth - \csllabelwidth}{\strut#1\strut}}

\usepackage{graphicx}
\usepackage{placeins}
\usepackage{caption}
\usepackage{bookmark}
\IfFileExists{xurl.sty}{\usepackage{xurl}}{} 
\makeatletter
\@ifundefined{xmpquote}{}{}
\makeatother
\hypersetup{
  colorlinks=true,
  linkcolor={Maroon},
  filecolor={Maroon},
  citecolor={Blue},
  urlcolor={Blue},
  pdfcreator={LaTeX via pandoc}}

\author{}
\date{}

\begin{document}

\title{Polymer Membrane Tensegrity: Inverse Design of Polymer Films Morphing into Freeform 3D Surfaces with Digital Photopatterning Technique}
\author{\parbox{\textwidth}{\centering Shuto Ito$^{1,*}$, Yuta Shimoda$^{1,2,3,*}$, Haruka Fukunishi$^{1}$, Mikihiro Hayashi$^{1,4}$\\ \normalsize $^{1}$ Biomatter Lab, Toyonaka, Osaka, Japan \\ $^{2}$ Graduate School of Arts and Sciences, The University of Tokyo, Tokyo, Japan \\ $^{3}$ School of Science and Technology, Meiji University, Kanagawa, Japan \\ $^{4}$ Department of Chemical Science and Engineering, School of Materials and Chemical Technology, Institute of Science Tokyo, 2-12-1 Ookayama, Meguro-ku, Tokyo 152-8550, Japan \\ $^{*}$ These authors contributed equally and are co-corresponding authors. \\ Correspondence: dr.shutoito@gmail.com}}
\date{}
\maketitle

\subsection{Abstract}\label{abstract}

In \emph{Metamorphosis of Plants} (1790), Goethe traced diverse plant
organs to transformations of a common leaf-like structure --- a
principle modern mechanics attributes to two material ingredients:
non-uniform in-plane strain from differential growth or shrinkage, and
spatially patterned stiffness. Here we translate this principle into a
synthetic fabrication framework called Polymer Membrane Tensegrity
(PMT). A flat elastomeric film swollen with a second monomer is
selectively UV-cured through a liquid-crystal display (LCD) photomask in
a single-side digital photopatterning step, producing rigid rods
embedded in a soft, shrinkable membrane. After the unreacted monomer is
extracted with a solvent and the film is dried, the membrane shrinks far
more than the rods, generating a \textasciitilde50\% in-plane strain
differential and a \textasciitilde2,000-fold modulus contrast ---
conditions under which the contracting membrane is held in tension by
mutually unconnected rods, a tensegrity-inspired arrangement within a
single film. An origami-based inverse design algorithm computes the rod
layout that morphs the film into a prescribed 3D surface. We demonstrate
PMT on a dome, a hyperbolic surface, and a gyroid unit cell ---
positively and negatively curved targets --- reproducing all three with
mean deviations of 1.0--2.1\% of the target size; perimeter curve
optimization halves the mean deviation of the gyroid. Because patterning
occurs on one side only, PMT eliminates the front-to-back alignment
demanded by bilayer methods, offering a scalable route from flat polymer
films to freeform 3D surfaces.

\subsection{Introduction}\label{introduction}

\begin{figure*}[!t]
\centering
\includegraphics[width=0.95\linewidth,height=\textheight,keepaspectratio,alt={Figure 1. Background. Polymer science (left) --- the four plant-organ morphologies that differential growth produces (twisting, helical twisting, saddle bending, and edge waving; top row) reproduced in polymer films by digital photopatterning (bottom row; technique from Fukunishi et al., 2023). Computational origami (right) --- a target 3D mesh is tessellated into a developable crease pattern by generalizing Resch's patterns (Tachi, 2013), and replacing its mountain creases with compression rods and its facets and valley creases with a pre-tensioned membrane yields a membrane tensegrity that settles into the target surface, realized here as a physical model (Shimoda et al., 2023).}]{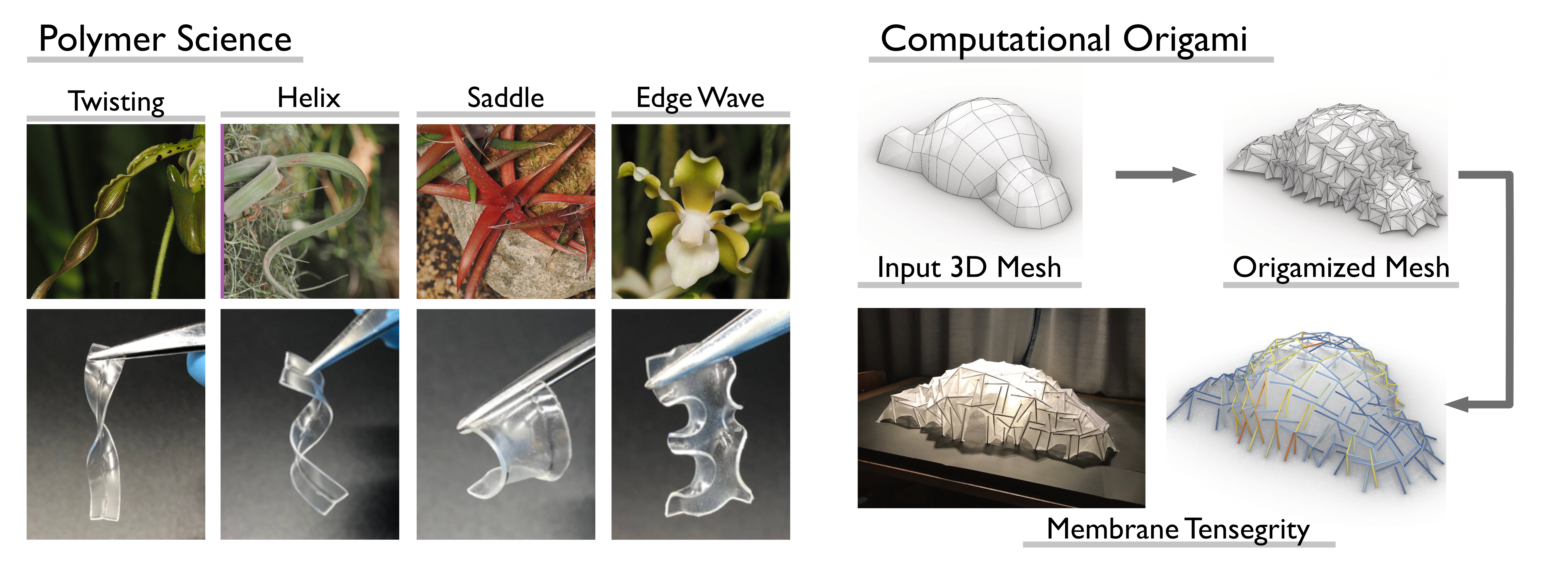}
\caption{\textbf{Figure 1.} Background. Polymer science (left) --- the
four plant-organ morphologies that differential growth produces
(twisting, helical twisting, saddle bending, and edge waving; top row)
reproduced in polymer films by digital photopatterning (bottom row;
technique from Fukunishi et al., 2023). Computational origami (right)
--- a target 3D mesh is tessellated into a developable crease pattern by
generalizing Resch's patterns (Tachi, 2013), and replacing its mountain
creases with compression rods and its facets and valley creases with a
pre-tensioned membrane yields a membrane tensegrity that settles into
the target surface, realized here as a physical model (Shimoda et al.,
2023).}
\end{figure*}

Goethe framed plant morphogenesis as repeated transformations of a
common leaf-like organ: sepals, petals, stamens, and fruits all derive
from modified leaves\textsuperscript{1}. Mechanics has since revealed
how each organ acquires its distinct 3D shape from a shared flat
precursor. The physical mechanism rests on two ingredients ---
non-uniform in-plane strain from differential growth or shrinkage, and
mechanical anisotropy imposed by the architecture of cellulose
microfibrils in cell walls and by macroscopic vein networks that stiffen
specific regions of the tissue\textsuperscript{2--4}. This dual
mechanism underlies a wide range of functional morphologies: Liang and
Mahadevan showed that differential edge growth drives the blooming of
lily petals\textsuperscript{5}, Huang et al.~unified four plant-organ
shapes into a morphological phase diagram governed by the magnitude and
distribution of growth strain\textsuperscript{6} (Figure 1, left), Armon
et al.~demonstrated that the chiral opening of seed pods is controlled
by the roughly ±45° orientation of cellulose fibers in bilayer
walls\textsuperscript{7}, and Guo et al.~showed theoretically that a
non-shrinking midvein constraining a uniformly shrinking leaf lamina ---
a natural rod-membrane system --- governs the curling and folding of
drying leaves\textsuperscript{8}. What unifies these systems is not the
sign of the deformation but the mismatch: neighbouring regions of one
thin organ acquire different in-plane strains --- because one grows
faster, swells more, or shrinks less than the other --- and the
resulting incompatibility prescribes a non-Euclidean target metric that
the tissue can only accommodate by leaving the plane. Whether the strain
differential is written as growth, as in a blooming petal, or as
shrinkage, as in a drying leaf, the mechanics is the same. It is this
principle, rather than any particular biological mechanism, that the
fabrication strategy pursued here reproduces in a synthetic film.

In engineered systems, differential-strain-based shape morphing has been
most extensively explored in hydrogels: Kim et al.~used halftone gel
lithography to prescribe non-Euclidean target metrics through spatially
graded cross-link density\textsuperscript{9}, and Gladman et
al.~achieved biomimetic 4D printing by aligning cellulose fibrils during
direct-ink writing to program anisotropic swelling\textsuperscript{10}.
Patterning of mechanical stiffness has likewise been used to direct
shape change, for example through rigid--soft polymer multilayers or
inkjet-printed UV-curable composites that fold into origami
structures\textsuperscript{11--13}. These studies collectively show that
both strain heterogeneity and stiffness heterogeneity --- the same two
ingredients that drive plant morphogenesis --- can be harnessed to
program 3D shapes from flat sheets. However, hydrogel-based approaches
require aqueous environments and produce transient shapes, while
existing polymer self-folding methods typically require patterning on
both sides of the sheet to define fold direction, necessitating
specialized equipment and precise front-to-back alignment.

In polymer systems, digital photopatterning provides a practical route
to embed such heterogeneity in a dry, self-standing film. In earlier
work from our group, Fukunishi et al.~showed that a parent cross-linked
film swollen with a second monomer can be irradiated area-selectively
using a liquid-crystal display (LCD) 3D printer, producing patterned
regions that differ in stiffness and in the strain they recover after
loading, and thereby enabling buckling and wrinkling responses in a flat
sheet\textsuperscript{14}. Kuwada et al.~further demonstrated that
multipolymeric patterning can realize tensegrity-inspired polymer films
with progressive bending stiffness, indicating that rod-membrane force
balance can be embodied in a thin polymer
architecture\textsuperscript{15}. In the classical tensegrity framework,
isolated rigid rods are stabilized by a network of tensile elements
without mutual contact between the rods, producing lightweight
structures with tunable stiffness\textsuperscript{16}. This structural
principle has been recognized across scales, from cellular
cytoskeletons\textsuperscript{17} to architectural structures, and
Kuwada et al.'s work showed that it can also be materialized at the
polymer-film level.

A remaining challenge is inverse design: given a target 3D surface, how
should the in-plane material pattern be computed? In computational
origami, Tachi showed that a single uncut sheet can be folded into any
given polyhedral surface (Origamizer)\textsuperscript{18,19}, and
established a framework for generating developable freeform variations
of sheet patterns and, by generalizing Resch's patterns, for
tessellating a target surface into a developable crease pattern (Figure
1, right)\textsuperscript{20,21}. We previously extended this line of
work to membrane tensegrity, showing that if the mountain creases of
such a tessellation are replaced by compression rods and the remaining
facets and creases by a membrane held under initial tension, the
assembly settles into the target surface\textsuperscript{22}. In
parallel, broader shape-programming studies have shown that inverse
design can be posed for active sheets through prescribed metrics and
local anisotropy, even for surfaces with complex
curvature\textsuperscript{23}.

Here, we combine these two previously separate directions, digital
photopatterning in polymer films and inverse-designed membrane
tensegrity, into a single fabrication framework that we call Polymer
Membrane Tensegrity (PMT). By using digital photopatterning to implement
the rod and membrane regions computed from an origami-based inverse
design algorithm, we seek to transform a planar polymer film into
freeform 3D surfaces while retaining a simple 2D fabrication process.
Unlike hydrogel shape-morphing, the resulting structures are dry,
permanent, and self-supporting in ambient conditions. The following
sections quantify the shrinkage differential and modulus contrast that
make this rod--membrane mechanism possible and then demonstrate morphing
into a dome, a hyperbolic surface, and a gyroid unit cell --- a
positively curved surface and two negatively curved ones --- to
illustrate that PMT can access freeform geometries.

\subsection{Materials and Methods}\label{materials-and-methods}

\begin{figure*}[!t]
\centering
\includegraphics[width=0.95\linewidth,height=\textheight,keepaspectratio,alt={Figure 2. Overview of the PMT pipeline. Top row (computational). Input 3D Mesh --- the target surface. Origamized Mesh --- the target converted into a Resch-type origami tessellation. Crease Pattern --- the creases extracted from that mesh. Irradiation Pattern --- the crease pattern rasterized to the LCD pixel grid as a binary image, in which white denotes the rod regions to be polymerized and black the membrane regions left unexposed. Bottom row (experimental). Immersion --- the cross-linked EA/tBA parent film is immersed in the MMA/TEGDMA monomer solution until equilibrium swelling. Area-selective Photopolymerization --- the monomer-absorbed film is sandwiched between PET sheets on the LCD screen and exposed to 405 nm UV through the irradiation pattern, polymerizing the second monomer only where the pattern is open. Area-selective Shrinkage \& Morphing --- after unreacted monomer is extracted in acetone and the film is dried, the non-irradiated membrane contracts while the irradiated rods retain their swollen dimensions, and the resulting in-plane strain differential drives the film into the target surface.}]{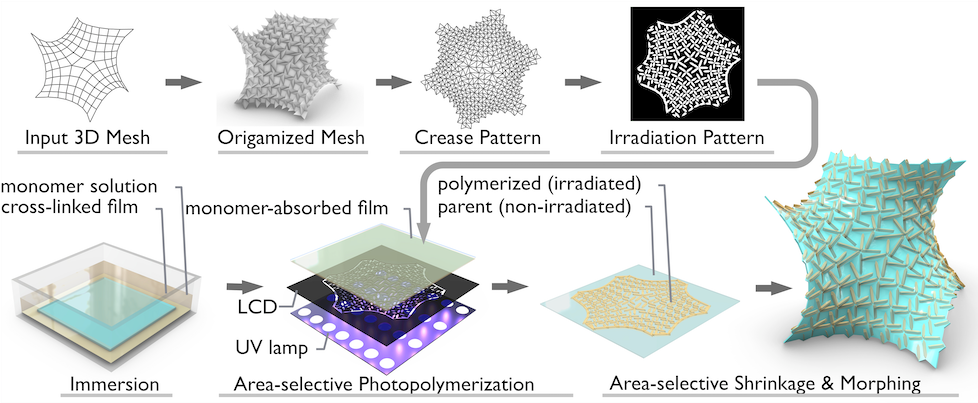}
\caption{\textbf{Figure 2.} Overview of the PMT pipeline. \textbf{Top
row (computational).} \emph{Input 3D Mesh} --- the target surface.
\emph{Origamized Mesh} --- the target converted into a Resch-type
origami tessellation. \emph{Crease Pattern} --- the creases extracted
from that mesh. \emph{Irradiation Pattern} --- the crease pattern
rasterized to the LCD pixel grid as a binary image, in which white
denotes the rod regions to be polymerized and black the membrane regions
left unexposed. \textbf{Bottom row (experimental).} \emph{Immersion} ---
the cross-linked EA/tBA parent film is immersed in the MMA/TEGDMA
monomer solution until equilibrium swelling. \emph{Area-selective
Photopolymerization} --- the monomer-absorbed film is sandwiched between
PET sheets on the LCD screen and exposed to 405 nm UV through the
irradiation pattern, polymerizing the second monomer only where the
pattern is open. \emph{Area-selective Shrinkage \& Morphing} --- after
unreacted monomer is extracted in acetone and the film is dried, the
non-irradiated membrane contracts while the irradiated rods retain their
swollen dimensions, and the resulting in-plane strain differential
drives the film into the target surface.}
\end{figure*}

\subsubsection{Material Formulation}\label{material-formulation}

\textbf{Parent film.} A cross-linked elastomeric film was prepared from
ethyl acrylate (EA) and \emph{tert}-butyl acrylate (tBA) as comonomers
(EA:tBA ≈ 44:56 by weight), with 1,4-butanediol diacrylate (BDA) as a
cross-linker and diphenyl(2,4,6-trimethylbenzoyl)phosphine oxide (TPO, 1
wt\%) as a photoinitiator. The monomer solution was poured into the
resin vat of an LCD 3D printer (Anycubic Photon Mono 2) and spread into
a thin layer, then flood-exposed with the printer's 405 nm UV LED to
cure the entire area at once. The cured film was peeled from the vat to
yield a flat, elastomeric parent film with a nominal thickness of 0.3
mm; the measured thickness and its swollen value are reported in
Results.

\textbf{Second monomer solution.} Methyl methacrylate (MMA) was used as
the second monomer with triethylene glycol dimethacrylate (TEGDMA) as a
cross-linker (MMA:TEGDMA ≈ 41:59 by weight) and TPO (1 wt\%) as a
photoinitiator. Upon polymerization, MMA/TEGDMA regions form a glassy,
rigid polymer with a glass transition temperature (T\(_\mathrm{g}\))
well above 100 °C, which serves as the ``rod'' component of the
tensegrity structure.

\subsubsection{Digital Photopatterning}\label{digital-photopatterning}

The parent film was immersed in the second monomer solution for ≥ 12 h,
until equilibrium swelling was confirmed by constant mass. The swollen
film was then sandwiched between PET films and placed on the LCD screen
of the same 3D printer, whose pixel pitch is 51 μm. A programmed
irradiation pattern, exported from the inverse design algorithm as a
bitmap image, was projected through the LCD for 80--90 s to selectively
polymerize the second monomer in the designated rod regions. This
exposure time was fixed in preliminary experiments as the shortest
exposure that polymerized the rod regions through the full thickness of
the film without overexposure. After irradiation, the film was removed
from the PET sandwich and immersed in acetone to extract all unreacted
monomer from the non-irradiated membrane regions. Because acetone swells
the network more strongly than the monomer solution itself, the film
dimensions during washing transiently exceed the swollen state (Table 1,
step 5). Upon removal from the acetone bath, the solvent evaporates and
the non-irradiated membrane shrinks substantially --- returning toward
the dimensions of the original parent film --- while the irradiated rod
regions retain their swollen dimensions. This differential shrinkage
produces the in-plane strain contrast that drives 2D-to-3D morphing.

\subsubsection{Inverse Design Algorithm}\label{inverse-design-algorithm}

The origami-based inverse design algorithm that we proposed
previously\textsuperscript{22} was implemented in Rhinoceros/Grasshopper
and applied to compute the irradiation patterns. The pipeline proceeds
as follows. A target 3D mesh is converted into an origami tessellation
by generalizing Resch's patterns\textsuperscript{21}, and the
tessellation is then optimized for developability using Crane (v0.2.4),
a Grasshopper implementation of Tachi's freeform origami
framework\textsuperscript{24}. The rod layout is not optimized as an
independent variable: the mountain creases of the resulting developable
tessellation are taken directly as the rod regions (irradiated areas),
and the remaining facets and creases become the shrinking membrane. The
binary irradiation pattern is exported from this assignment.

To this pipeline we add a perimeter curve treatment (Figure 4). The need
for it becomes clear when the length of the outline before and after
shrinkage is compared with the length the target requires. Let \(l_1\)
be the length of a boundary segment of the flat irradiation pattern,
\(l_2\) the length of the same segment after shrinkage, \(l_\mathrm{t}\)
the length of the corresponding segment of the target outline, \(s\) the
shrink ratio of the membrane, and \(k\) the fraction of the boundary
segment occupied by rod. The three boundary conditions differ in how
\(l_2\) is set. If the pattern is taken directly from the tessellation
with no constraint placed on the outline (``no perimeter''), the
boundary consists of membrane only and shrinks freely,

\[l_2 = s\, l_1 \neq l_\mathrm{t}; \tag{1}\]

the inverse problem is simply not solved along the outline. In the
partial perimeter design, rod segments are inserted along the outline;
because the rod retains its dimensions while the membrane contracts,

\[l_2 = \bigl[s\,(1-k) + k\bigr]\, l_1 = l_\mathrm{t}, \tag{2}\]

with \(k\) chosen so that the equality holds. The perimeter length is
then correct, but the boundary rods remain disconnected and the fold
angles of the boundary creases are unconstrained. In the full perimeter
design the tessellation is generated with the fold angles along the
outline constrained to zero, so that the boundary segments of the
origamized mesh coincide with the target outline,
\(l_1 = l_\mathrm{t}\); Eq. (2) is then satisfied only by \(k = 1\),
that is, the boundary rods meet end to end and form a single continuous
locking ring around the perimeter,

\[l_2 = l_1 = l_\mathrm{t}, \tag{3}\]

which holds the edge flat and locks its length to that of the target.
The constraint \(l_1 = l_\mathrm{t}\) is not imposed afterwards but
solved together with the interior: the zero fold angle along the outline
enters the same Crane optimization as the developability and
target-fitting constraints of Tachi's framework\textsuperscript{24}, so
that the interior tessellation and the boundary ring are found in one
solve. This boundary treatment is not part of the original algorithm of
Shimoda et al., which solves only for the interior of the tessellation.

\subsubsection{Shape Evaluation}\label{shape-evaluation}

The morphed films were fixed in their 3D configuration by secondary UV
curing (5 min under a 405 nm flood lamp at \textasciitilde20 mW
cm\(^{-2}\)) and scanned using a structured-light 3D scanner (MINI 2,
Revopoint). The scanned meshes were exported as STL files and aligned to
the target meshes by rigid iterative-closest-point (ICP) registration,
initialized from principal-axis alignment (four sign hypotheses) and
retaining the converged pose with the lowest mean point-to-surface
distance. For every scan vertex, the deviation was computed as the
distance to the closest point on the target mesh and normalized by the
bounding-box diagonal of the target mesh; deviation maps are color-coded
from 0\% (blue) to 5\% (red) (Figure 3 and Supplementary Figure S1).
Reporting the deviation as a fraction of the target size rather than as
an absolute distance follows the recommendation of Wang and Chortos for
shape-morphing devices, whose accuracy should be compared relative to
the magnitude of the target deformation\textsuperscript{25}; the
bounding-box diagonal is used here so that the three targets, whose
out-of-plane extents differ, are compared on a common footing.
Deviations are evaluated over the scanned surface; for the dome, the
scan does not cover the lowermost rim band. To characterize how the
deviation is distributed over the surface, the mean deviation was
additionally evaluated in 2 mm shells of distance from the free edge of
the target mesh; per-specimen deviation statistics are summarized in
Supplementary Table S1, and the deviation-versus-distance profiles of
the hyperbolic and gyroid specimens are given in Supplementary Figure
S2. The ICP alignment is validated against an independent
evolutionary-solver alignment in Supplementary Note S1 and Supplementary
Table S2.

\subsubsection{Mechanical Testing}\label{mechanical-testing}

The Young's modulus and tensile strength of the rod (irradiated,
MMA/TEGDMA-polymerized) and membrane (non-irradiated, EA/tBA parent
film) materials were characterized by uniaxial tensile testing using a
universal testing machine (Autograph AGS-X, Shimadzu) at a crosshead
speed of 5 mm/min. Flat plate specimens were used with a grip distance
of 15 mm: 30 mm wide and 0.50 mm thick for the rod material (n = 4) and
20 mm wide and 0.30 mm thick for the membrane material (n = 5). Values
are reported as mean ± SD.

\subsection{Results and Discussion}\label{results-and-discussion}

\subsubsection{Shrinkage and Mechanical
Properties}\label{shrinkage-and-mechanical-properties}

The parent films were 0.33 ± 0.02 mm thick (n = 10). The in-plane
dimensional changes of the parent films were measured at each processing
step (Table 1; n = 20 per group, except step 5: n = 9). The irradiated
(irr) regions, where the second monomer was polymerized, retained their
swollen dimensions, while the non-irradiated (non) regions shrank back
upon acetone washing and drying.

\begin{table*}[!t]
\centering\small
\caption{\textbf{Table 1.} Dimensional changes of the parent films
through the processing steps.}
\begin{tabular}{@{}
  >{\raggedright\arraybackslash}p{(\linewidth - 6\tabcolsep) * \real{0.0272}}
  >{\raggedright\arraybackslash}p{(\linewidth - 6\tabcolsep) * \real{0.8095}}
  >{\raggedright\arraybackslash}p{(\linewidth - 6\tabcolsep) * \real{0.0816}}
  >{\raggedright\arraybackslash}p{(\linewidth - 6\tabcolsep) * \real{0.0816}}@{}}
\toprule\noalign{}
\begin{minipage}[b]{\linewidth}\raggedright
Step
\end{minipage} & \begin{minipage}[b]{\linewidth}\raggedright
Description
\end{minipage} & \begin{minipage}[b]{\linewidth}\raggedright
irr (mm)
\end{minipage} & \begin{minipage}[b]{\linewidth}\raggedright
non (mm)
\end{minipage} \\
\midrule\noalign{}
1 & Initial (2×2 cm cut) & 20.3 ± 0.4 & 19.8 ± 0.5 \\
2 & After second monomer immersion & 34.3 ± 0.7 & 33.8 ± 0.7 \\
3 & Before UV irradiation (in PET sandwich) & 34.3 ± 0.8 & --- \\
4 & After UV irradiation & 33.6 ± 0.8 & --- \\
5 & After acetone washing & 36.5 ± 0.8 & 37.3 ± 1.2 \\
6 & After drying & 33.0 ± 0.7 & 21.8 ± 0.6 \\
\bottomrule\noalign{}
\end{tabular}
\end{table*}

\begin{table*}[!t]
\centering\small
\caption{\textbf{Table 2.} Key dimensional ratios and in-plane strain
differential.}
\begin{tabular}{@{}
  >{\raggedright\arraybackslash}p{(\linewidth - 2\tabcolsep) * \real{0.5000}}
  >{\raggedright\arraybackslash}p{(\linewidth - 2\tabcolsep) * \real{0.5000}}@{}}
\toprule\noalign{}
\begin{minipage}[b]{\linewidth}\raggedright
Ratio
\end{minipage} & \begin{minipage}[b]{\linewidth}\raggedright
Value
\end{minipage} \\
\midrule\noalign{}
Swelling ratio (step 2 / step 1) & 1.70 (both irr and non) \\
Shrinkage ratio, irr (step 6 / step 2) & 0.963 \\
Shrinkage ratio, non (step 6 / step 2) & 0.644 \\
\textbf{In-plane strain differential (ratio of retained dimensions,
0.963/0.644 − 1)} & \textbf{0.50 (50\%)} \\
\bottomrule\noalign{}
\end{tabular}
\end{table*}

Upon drying, the non-irradiated membrane regions shrink to 64.4\% of the
swollen dimensions, while the irradiated rod regions retain 96.3\%,
producing a 50\% in-plane strain differential between the two material
phases (0.963/0.644 = 1.50; Table 2). This large differential strain
drives the 2D-to-3D morphing transformation, where the membrane's
contractile stress is held by the embedded rigid rods.

\begin{table*}[!t]
\centering\small
\caption{\textbf{Table 3.} Mechanical properties of rod and membrane
materials (mean ± SD; n = 4 rod and 5 membrane specimens; specimen
geometry in Methods).}
\begin{tabular}{@{}
  >{\raggedright\arraybackslash}p{(\linewidth - 4\tabcolsep) * \real{0.3500}}
  >{\raggedright\arraybackslash}p{(\linewidth - 4\tabcolsep) * \real{0.3500}}
  >{\raggedright\arraybackslash}p{(\linewidth - 4\tabcolsep) * \real{0.3000}}@{}}
\toprule\noalign{}
\begin{minipage}[b]{\linewidth}\raggedright
Property
\end{minipage} & \begin{minipage}[b]{\linewidth}\raggedright
Rod (MMA/TEGDMA-polymerized)
\end{minipage} & \begin{minipage}[b]{\linewidth}\raggedright
Membrane (EA/tBA parent)
\end{minipage} \\
\midrule\noalign{}
Young's modulus \emph{E} & 0.77 ± 0.16 GPa & 0.38 ± 0.05 MPa \\
Tensile strength σ\_max (MPa) & 13 ± 9 & 0.53 ± 0.12 \\
\bottomrule\noalign{}
\end{tabular}
\end{table*}

Uniaxial tensile testing gave a Young's modulus of 0.77 ± 0.16 GPa for
the rod material and 0.38 ± 0.05 MPa for the membrane (Table 3;
stress--strain curves of all specimens in Supplementary Figure S3). This
\textasciitilde2,000-fold difference in elastic modulus ensures that the
irradiated regions act as rigid structural elements, while the
non-irradiated membrane regions accommodate the shrinkage-induced
deformation, consistent with the intended division of roles between rods
and membrane.

\subsubsection{A Tensegrity-Inspired Rod--Membrane
Arrangement}\label{a-tensegrity-inspired-rodmembrane-arrangement}

The name Polymer Membrane Tensegrity refers to the arrangement of the
two phases rather than to a claim about the internal force state. In
classical tensegrity, isolated compression members are stabilized by a
tension network without contact between them\textsuperscript{16}; in PMT
the MMA/TEGDMA-polymerized rod regions likewise do not form a connected
framework but are individually embedded in the contracting EA/tBA
membrane, whose shrinkage to 64.4\% of its swollen dimension against
rods that retain 96.3\% places the membrane in tension and the rods
under compression, and it is this stress state that holds the 3D shape
after drying. The analogy stops there. Unlike the struts of a classical
tensegrity, the rods are bonded to the membrane along their entire
length, so their bending and buckling are governed by the rod--membrane
composite --- the membrane restrains the rods, and the rods in turn are
what the membrane's contraction acts against --- rather than by the rods
alone; the \textasciitilde2,000-fold modulus contrast (Table 3) makes
this division of roles possible but is not by itself a criterion for
tensegrity. We do not measure the internal force balance directly. Two
observations from our own earlier work nevertheless support the picture:
at the architectural scale, membrane-tensegrity structures based on the
same origami tessellations stand as self-supporting
assemblies\textsuperscript{22}, and in patterned polymer films
fabricated by the same photopatterning route, Kuwada et al.~observed
progressive bending stiffening arising from the membrane tension
generated by the protruding rods\textsuperscript{15}. The permanent,
self-supporting shapes reported here are consistent with the same
rod--membrane stress state. Tensegrity-inspired materials have recently
been demonstrated by other routes as well --- Xue et al.~prestress a
hydrogel network by growing tyrosine crystals in
situ\textsuperscript{26} --- and PMT differs from these in obtaining the
prestress from geometric patterning and differential shrinkage, in a dry
film with a permanent shape.

\subsubsection{3D Surface Morphing}\label{d-surface-morphing}

To demonstrate that PMT can produce freeform surfaces, we selected three
targets in order of increasing demand on the control of the boundary: a
dome, with positive Gaussian curvature; a hyperbolic surface, with
negative Gaussian curvature; and a gyroid unit cell, a patch of the
triply periodic minimal surface, as a second negatively curved target of
more complex shape. This order reflects the mechanics of a shrinking
tessellation. An unconstrained edge contracts more readily than the
interior, so a film left to itself tends toward positive curvature, and
a dome places the least demand on the boundary; negative curvature
instead requires the edges to remain longer than the interior would make
them, and therefore calls for a more careful treatment of the boundary,
which the gyroid, with its longer and more intricate outline, tests
further. For each target, the irradiation pattern computed by the
inverse design algorithm was projected onto the swollen parent film,
which self-morphed into the target 3D surface upon acetone washing and
drying (Figure 3). All three specimens were fabricated with the partial
perimeter design, in which rod segments along the outline match the
perimeter length after shrinkage to that of the target (Eq. 2; Methods);
the boundary treatment itself is compared systematically below.

The dome target (K \textgreater{} 0) produced a smooth, positively
curved surface that closely matched the input mesh: the mean deviation
over the scanned surface is 0.61 mm, corresponding to 1.0\% of the
bounding-box diagonal of the target (maximum 2.3 mm). The hyperbolic
target (K \textless{} 0) was reproduced at a mean deviation of 1.20 mm
(2.1\%; maximum 6.8 mm); the largest deviations concentrate at two
opposite free corners, where the saddle curvature imposes the highest
bending demands. The gyroid unit cell self-morphed at a mean deviation
of 1.00 mm (1.7\%; maximum 3.7 mm), an accuracy comparable to that of
the dome and hyperbolic targets. Across all three targets, the interior
of the surface reproduces the target to within roughly 1 mm. How the
remaining deviation is distributed differs between the targets
(Supplementary Figure S2): for the hyperbolic target the deviation peaks
a few millimetres inside the free edge --- reaching about 2.8\% at 4--6
mm --- and decays to below 1\% toward the interior, whereas for the
gyroid the deviation is not localized at the edge and is largest 10--16
mm inside it (the dome scan does not cover the rim; Methods).
Edge-localized error of this kind is the failure mode that the perimeter
treatment addresses, as discussed below.

Across all three targets, the fabricated films spontaneously adopted 3D
configurations that are qualitatively faithful to the input meshes,
confirming that the strain differential and modulus contrast reported
above (Tables 2 and 3) provide a sufficient mechanical driving force for
morphing into surfaces ranging from positive to spatially varying
negative curvature.

\begin{figure*}[!t]
\centering
\includegraphics[width=0.95\linewidth,height=\textheight,keepaspectratio,alt={Figure 3. Inverse design and fabrication of the three target surfaces. From left to right: input 3D mesh, origamized mesh, irradiation pattern, and deformed film with deviation map. Top: Dome (positive Gaussian curvature). Middle: Hyperbolic (negative Gaussian curvature). Bottom: Gyroid unit cell (negative Gaussian curvature). All three specimens use the partial perimeter design (Methods).}]{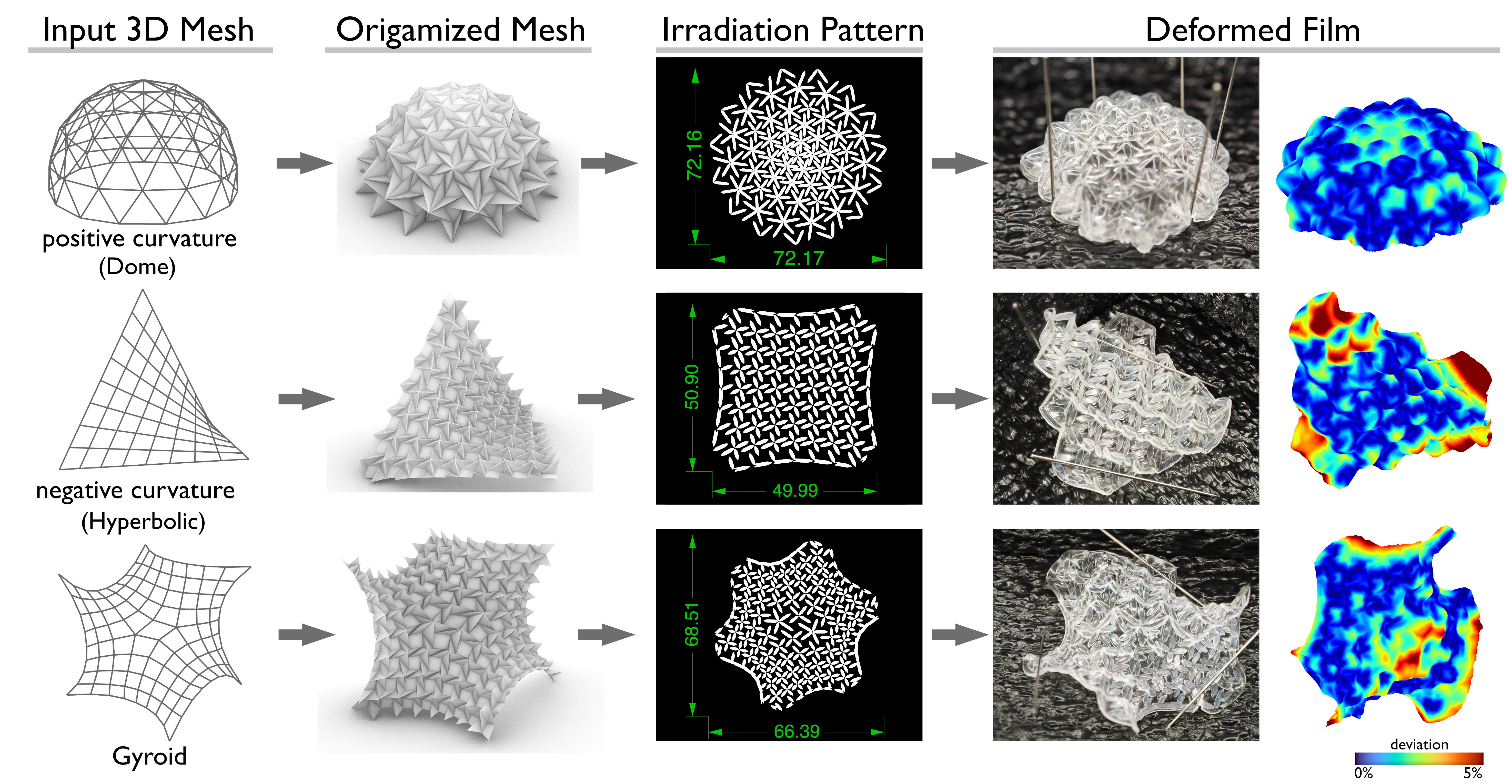}
\caption{\textbf{Figure 3.} Inverse design and fabrication of the three
target surfaces. From left to right: input 3D mesh, origamized mesh,
irradiation pattern, and deformed film with deviation map. Top: Dome
(positive Gaussian curvature). Middle: Hyperbolic (negative Gaussian
curvature). Bottom: Gyroid unit cell (negative Gaussian curvature). All
three specimens use the partial perimeter design (Methods).}
\end{figure*}

\subsubsection{Perimeter Curve
Optimization}\label{perimeter-curve-optimization}

The accuracy of the morphed 3D surface depends critically on how the
boundary is treated in the inverse design algorithm. We compared three
boundary conditions --- no perimeter constraint, matching of the
perimeter length alone, and the full perimeter treatment in which the
boundary fold angles are additionally constrained to zero (Methods, Eqs.
1--3 and Figure 4) --- using the gyroid unit cell (Figure 4), the target
on which the boundary matters most: it is negatively curved, so its edge
must be held longer than a free edge would contract to, and its outline
is long.

Without a perimeter curve (``no perimeter''), the inverse problem is
solved only for the interior of the mesh. The resulting film morphs into
a shape that qualitatively resembles the target but exhibits systematic
dimensional mismatch at the boundary: the perimeter of the morphed film
does not match the target perimeter length, causing the edges to curl
inward or flare outward uncontrollably. Averaged over the surface, the
deviation from the target mesh is 3.0\% of the bounding-box diagonal
(1.79 mm on average), the largest of the three conditions.

With a partial perimeter curve, rod segments are inserted along the
outline so that the perimeter length after shrinkage matches the target
(Eq. 2). While this eliminates the gross dimensional mismatch, the
boundary remains mechanically underconstrained --- the perimeter rods
lack the fold-angle optimization that stabilizes interior facets. As a
result, the edge exhibits visible waviness and local buckling,
indicating that perimeter length matching alone is insufficient for
structural stability. The mean deviation falls to 1.8\% (1.10 mm), an
improvement over the no-perimeter condition but still short of what
boundary rods with optimized fold angles achieve. The gyroid specimen of
Figure 3, fabricated independently with the same partial perimeter
design, gave a closely similar mean deviation (1.7\%; 1.00 mm) and a
similar deviation-versus-distance profile (Supplementary Figure S4),
indicating that the accuracy of this design is reproducible.

With the full perimeter curve design, the origami algorithm constrains
the fold angles of the boundary creases to zero, so that the boundary
rods close into a continuous ring whose length equals the target
perimeter (Eq. 3) and the edges are held flat. This produces a morphed
surface with stable, well-defined edges that closely follow the target
geometry, and the mean deviation drops further to 1.3\% (0.79 mm) --- a
56\% reduction relative to the no-perimeter condition. The three
conditions therefore rank monotonically in accuracy (3.0\%, 1.8\%,
1.3\%), and the improvement is obtained even though the full perimeter
design uses a smaller irradiation pattern (59.34 × 58.95 mm) than the
other two (68.51 × 66.39 mm), because the perimeter length is matched to
the target rather than left to shrink freely. These results demonstrate
that explicit boundary optimization is essential for high-fidelity shape
morphing. The reason, discussed below, is that an unconstrained outline
folds where it should not, and the locking ring removes that freedom.

\begin{figure*}[!t]
\centering
\includegraphics[width=0.95\linewidth,height=\textheight,keepaspectratio,alt={Figure 4. Effect of perimeter curve design on morphing accuracy. For each condition, top: irradiation pattern (dimensions in mm) and morphed film; bottom: close-up of the boundary in the irradiation pattern, the target origami, and the result, with the length budget of Eqs. 1--3 (l\_1, boundary segment of the flat pattern; l\_2, the same segment after shrinkage; l\_\textbackslash mathrm\{t\}, corresponding segment of the target outline; s, shrink ratio of the membrane; k, fraction of the segment occupied by rod). Left: no perimeter curve --- the inverse problem is not solved at the boundary (Eq. 1), resulting in dimensional mismatch. Center: partial perimeter curve --- rod segments make the perimeter length fit (Eq. 2) but the boundary is unconstrained in fold angle, so the structure is unstable and wavy. Right: full perimeter curve with origami optimization --- the boundary fold angles are set to zero so that l\_1 = l\_\textbackslash mathrm\{t\} and the boundary rods form a continuous ring (Eq. 3); the perimeter length fits and the structure is stable. Deviation maps of the three specimens are shown in Supplementary Figure S1.}]{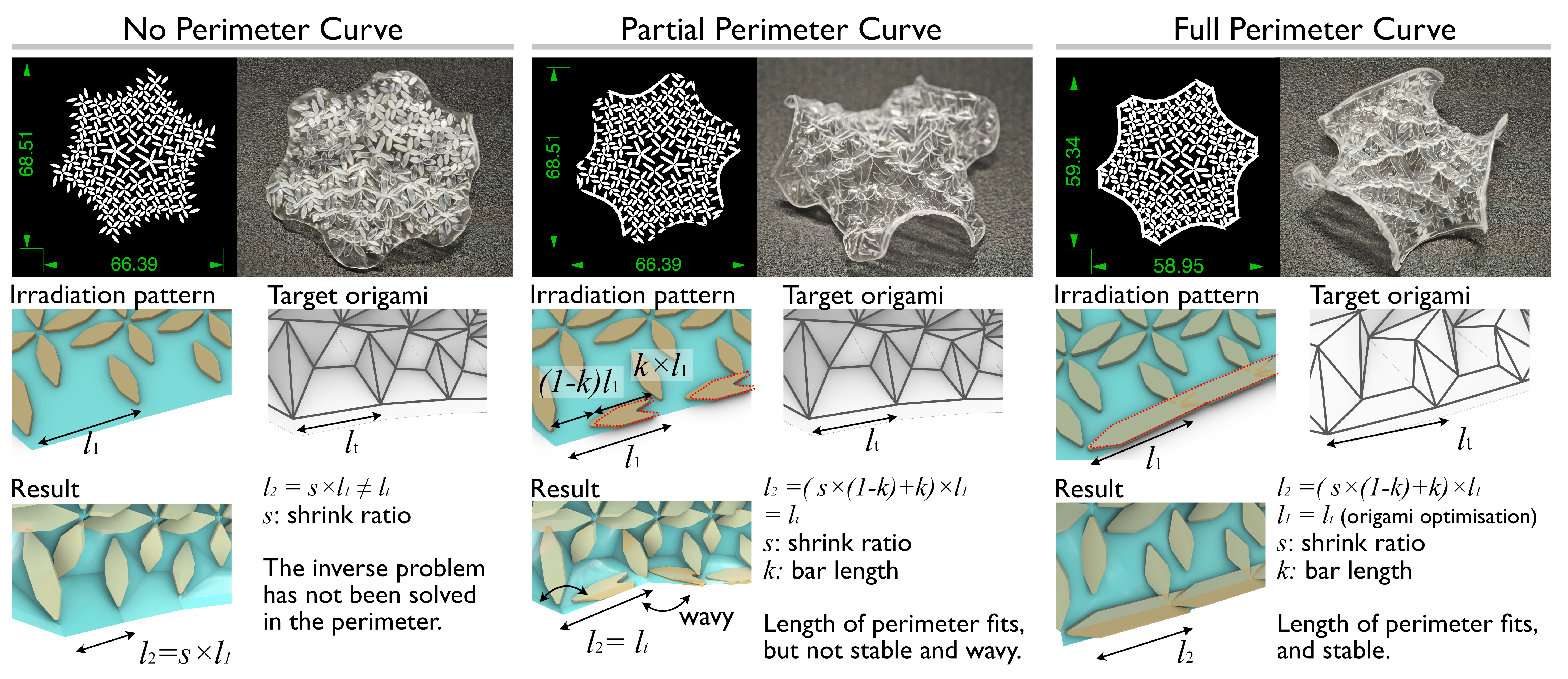}
\caption{\textbf{Figure 4.} Effect of perimeter curve design on morphing
accuracy. For each condition, top: irradiation pattern (dimensions in
mm) and morphed film; bottom: close-up of the boundary in the
irradiation pattern, the target origami, and the result, with the length
budget of Eqs. 1--3 (\(l_1\), boundary segment of the flat pattern;
\(l_2\), the same segment after shrinkage; \(l_\mathrm{t}\),
corresponding segment of the target outline; \(s\), shrink ratio of the
membrane; \(k\), fraction of the segment occupied by rod). Left: no
perimeter curve --- the inverse problem is not solved at the boundary
(Eq. 1), resulting in dimensional mismatch. Center: partial perimeter
curve --- rod segments make the perimeter length fit (Eq. 2) but the
boundary is unconstrained in fold angle, so the structure is unstable
and wavy. Right: full perimeter curve with origami optimization --- the
boundary fold angles are set to zero so that \(l_1 = l_\mathrm{t}\) and
the boundary rods form a continuous ring (Eq. 3); the perimeter length
fits and the structure is stable. Deviation maps of the three specimens
are shown in Supplementary Figure S1.}
\end{figure*}

\subsubsection{Boundary Design and the Perimeter
Curve}\label{boundary-design-and-the-perimeter-curve}

Our results show that the treatment of the boundary decides the accuracy
of the morphed surface (Figure 4). The reason is visible in the
specimens themselves. Creases in the interior of the tessellation are
held on every side by neighbouring facets, whereas creases that reach
the free edge are held on one side only; they fold more readily and
reach larger angles than interior creases, so the periphery of an
unconstrained film contracts more than its centre and unwanted folds
form along the outline, which curls inward or flares outward. A
positively curved target tolerates this tendency, because its edge is
meant to turn inward; a negatively curved target does not, because its
edge must remain longer than the interior would make it, and for such
targets the unwanted folds along the outline are the dominant source of
error. The perimeter treatment was designed to remove this failure mode
at its source. Setting the fold angle along the outline to zero in the
inverse design (Eq. 3) closes the boundary rods into a continuous
locking ring that matches the target perimeter and cannot fold, so that
the edge is held flat at the correct length while the interior is left
free to adopt the target. Matching the perimeter length alone (Eq. 2) is
not sufficient, because disconnected boundary rods still allow the
membrane between them to fold --- the waviness of the partial condition.
The deviation maps of the three specimens (Supplementary Figure S1) show
the same progression: without a perimeter curve the deviation along the
outline locally exceeds 5\% of the bounding-box diagonal, whereas with
the full perimeter design only thin strips along the free edges remain
above 3\% and the interior lies within about 1\%. The corresponding
deviation-versus-distance profiles (Supplementary Figure S4) show the
same contrast: the sharp edge-region peak of the no-perimeter specimen
is absent from the full-perimeter profile.

\subsubsection{Inverse Design: From Target Surface to Flat
Pattern}\label{inverse-design-from-target-surface-to-flat-pattern}

The inverse design problem --- computing a 2D material pattern that
morphs into a desired 3D surface --- has been approached from several
directions. Aharoni et al.~demonstrated universal inverse design for
nematic elastomer sheets by computing director fields that encode a
target metric --- from which the desired Gaussian curvature follows ---
supplemented, for geometrically demanding targets, by a prescribed
reference curvature imposed through a director gradient across the
thickness\textsuperscript{23}. Guseinov et al.~solved the inverse
problem for CurveUps using rigid tile assemblies that deploy into target
shapes upon release of prestress\textsuperscript{27}. Panetta et
al.~addressed surface-based inflatables through computational
optimization of seam patterns\textsuperscript{28}.

PMT's inverse design, based on the origami-tessellation algorithm of our
earlier work\textsuperscript{22}, takes a different path: it discretizes
the target surface into a developable origami tessellation and assigns
the rods to its mountain creases, so that the pattern is geometrically
compatible with the target by construction rather than by optimization.
This approach inherits the geometric rigor of computational
origami\textsuperscript{19,20} while embedding the result in a
continuous polymer film rather than a folded sheet.

Geometric compatibility alone, however, does not guarantee that the
target is reached. Shimoda et al.~evaluated the equilibrium shape of
their structures by dynamic relaxation and found that the shrinking
membrane pulls the assembly away from the target, which they corrected
by iteratively updating the membrane edge lengths while holding the rod
lengths fixed; they further found that star vertices in regions where
the target surface curves away from the rods can snap through and
invert, and reported the mean curvature of the target surface as an
empirical index for when this occurs --- in their normalized tests,
vertices on surfaces of small mean curvature inverted, whereas those on
more strongly curved surfaces did not\textsuperscript{22}. Both effects
are relevant to PMT. The second is expected to matter most where the
mean curvature is small, and the gyroid unit cell is the limiting case:
as a minimal surface its mean curvature vanishes everywhere, so no
vertex of its tessellation is stabilized against inversion by the
curvature of the target.

\subsubsection{Advantages of PMT}\label{advantages-of-pmt}

Several strategies transform flat sheets into programmed 3D surfaces,
and a quantitative comparison locates the niche that PMT occupies.
Hydrogel systems prescribe non-Euclidean target metrics through
spatially modulated swelling --- halftone gel lithography reaches linear
strains of \textasciitilde50--190\%\textsuperscript{9} and
direct-ink-written composites
\textasciitilde11--47\%\textsuperscript{10} --- but require aqueous
environments and produce transient shapes. Origami- and kirigami-based
metamaterials\textsuperscript{29,30} reach doubly curved targets only
through facet bending and snap-through and depend on fabricated hinges
and external actuation, and multi-material 4D
printing\textsuperscript{31,32} relies on sustained external stimuli.
Self-folding trilayers\textsuperscript{11} use a
\textasciitilde5,000-fold modulus contrast, of the same order as PMT's
\textasciitilde2,000-fold, but stack the stiff and soft phases through
the thickness, so that swelling is rectified into bending and the fold
direction must be encoded by front-to-back asymmetry. PMT instead places
rod and membrane side by side within a single layer: the
\textasciitilde50\% strain differential is obtained in a dry, permanent
film, and the shape change originates from an in-plane metric change
held in equilibrium by the embedded rods.

This in-plane arrangement carries a practical advantage: the entire
transformation is encoded in a single-side irradiation pattern. Systems
that rectify swelling into bending must break front-to-back symmetry ---
by double-sided micropatterning\textsuperscript{11}, face-specific
inkjet deposition\textsuperscript{12}, or assembly of separately
patterned modules\textsuperscript{13} --- demanding registration that
becomes harder as the film grows. In PMT the fold direction does not
rely on face asymmetry, so a consumer-grade LCD 3D printer suffices, and
the process is in principle compatible with roll-to-roll fabrication: a
continuous film fed through a swelling bath, irradiated from one side,
washed, and dried.

The same two ingredients that PMT exploits --- differential strain and
differential rigidity --- underlie functional morphologies throughout
the plant kingdom, where stiff tissues paired with shrinking ones
generate autonomous movements without metabolic
input\textsuperscript{4,33}. In seed pods, wheat awns, and the Venus
flytrap the two phases are stacked through the thickness and the
mismatch is rectified into bending, with snap-through where the organ is
doubly curved\textsuperscript{7,34}; in a drying leaf the non-shrinking
midvein instead lies beside the shrinking lamina and acts within the
plane of the organ\textsuperscript{8}. PMT adopts this in-plane
arrangement: the rods play the role of the midvein or of a stiffened
vein network, the membrane that of the contracting lamina, and digital
photopatterning prescribes their layout as differential cell expansion
does in the leaf.

\subsubsection{Outlook}\label{outlook}

Two refinements follow directly from the present limitations. First, the
process does not control the local fold angle: the shrink ratio of the
membrane is spatially uniform, so nothing in the pattern prescribes how
far each crease folds, whereas the origamized mesh of a target generally
calls for a distribution of fold angles; grayscale modulation of the
exposure could provide this control and raise the attainable accuracy.
Second, star vertices can invert by snap-through during drying, and
biasing each vertex toward its intended branch would make the process
robust for targets with many small spikes. Beyond these refinements,
replacing the membrane with a stimulus-responsive polymer would make the
transformation reversible, so that flat films could be stored compactly
and deploy into load-bearing surfaces on a simple trigger such as
humidity, and the inverse design algorithm extends naturally to
biomimetic functional geometries such as the wing-like surface that
\emph{Firmiana} forms from its flat fruit wall through unfolding and
reverse bending of the pericarp\textsuperscript{35}. In both directions
the complexity resides in a digital 2D pattern rather than in
fabrication hardware --- echoing Goethe's insight that diverse forms
arise from transformations of a common planar precursor.

\subsection{Conclusion}\label{conclusion}

We have introduced Polymer Membrane Tensegrity (PMT), a framework that
turns a flat elastomeric film into a prescribed 3D surface by patterning
rigid rods and a shrinking membrane side by side in a single layer.
Swelling a cross-linked EA/tBA film with a second monomer and curing it
area-selectively through an LCD photomask produces a \textasciitilde50\%
in-plane strain differential and a \textasciitilde2,000-fold modulus
contrast between the two phases --- so that the contracting membrane is
held in tension by mutually unconnected embedded rods, in a
tensegrity-inspired arrangement. An origami-based inverse design
algorithm computes the rod layout for a given target, and we
demonstrated morphing into a dome, a hyperbolic surface, and a gyroid
unit cell --- a positively curved surface and two negatively curved
ones. Explicit optimization of the perimeter curve proved essential:
without it, the boundary is left unsolved and the edges curl or flare,
whereas full perimeter optimization yields stable, well-defined edges.

Two features distinguish PMT from existing shape-morphing routes. First,
the shape change arises from an in-plane metric change rather than
through-thickness bending, so the entire transformation is encoded in a
single-side irradiation pattern and no front-to-back registration is
required. Second, the resulting structures are dry, permanent, and
self-supporting in ambient conditions, unlike hydrogel systems that hold
their shape only in solvent. Because the design complexity resides in a
digital bitmap rather than in the fabrication hardware --- a
consumer-grade LCD 3D printer suffices --- the process is in principle
compatible with roll-to-roll manufacturing. Replacing the membrane phase
with a stimulus-responsive polymer would make the transformation
reversible, opening a route to deployable and biomimetic functional
surfaces designed entirely in software.

\subsection{Acknowledgements}\label{acknowledgements}

This work was supported by JSPS KAKENHI Grant Number JP26KJ1027.

\subsection{Author Contributions}\label{author-contributions}

S.I. and Y.S. conceived the study. S.I. developed the experimental
system and performed sample fabrication and mechanical characterization.
Y.S. developed and implemented the inverse design algorithm. H.F. and
M.H. designed the materials. S.I. and Y.S. wrote the manuscript, and all
authors reviewed and approved the final version.

\subsection{Data Availability}\label{data-availability}

The data that support the findings of this study are available from the
corresponding authors upon reasonable request.

\subsection{Competing Interests}\label{competing-interests}

The authors declare no competing interests.

\FloatBarrier
\section*{References}\label{bibliography}
\addcontentsline{toc}{section}{References}

\protect\FloatBarrier
\phantomsection\label{refs}

\end{document}